\documentstyle[12pt,amssym]{article}
\def\case#1#2{{\textstyle{#1\over #2}}}
\def\cosech{\mathop{\rm cosech}\nolimits}
\def\sech{\mathop{\rm sech}\nolimits}
\def\C{\mbox{$\Bbb C$}}
\title{
GENERALIZED CONTINUITY EQUATION AND MODIFIED NORMALIZATION IN PT-SYMMETRIC
QUANTUM MECHANICS}

\author{B.\ BAGCHI\thanks{E-mail: bbagchi@cucc.ernet.in} \\
{\small \sl Department of Applied Mathematics, University of Calcutta,} \\
{\small \sl 92 Acharya Prafulla Chandra Road, Calcutta 700 009, India}\\[10pt] 
C. QUESNE\thanks{Directeur de recherches FNRS; E-mail: cquesne@ulb.ac.be} \\
{\small \sl Physique Nucl\'eaire Th\'eorique et Physique Math\'ematique,}\\ 
{\small \sl Universit\'e Libre de Bruxelles, Campus de la Plaine CP229,} \\ {\small \sl
Boulevard~du Triomphe, B-1050 Brussels, Belgium} \\[10pt] 
M. ZNOJIL\thanks{E-mail: znojil@ujf.cas.cz} \\
{\small \sl $^c$ \'{U}stav jadern\'e fyziky AV \v{C}R, 250 68 \v{R}e\v{z}, Czech
Republic}}
\date{ }
\begin{document}
\baselineskip=22pt plus 1pt minus 1pt
\maketitle

\begin{abstract}
The continuity equation relating the change in time of the position probability density to
the gradient of the probability current density is generalized to PT-symmetric quantum
mechanics. The normalization condition of eigenfunctions is modified in accordance with
this new conservation law and illustrated with some detailed examples.
\end{abstract}

\vspace{1cm}
\noindent
Running head: Generalized Continuity Equation
\newpage
%
%
\section{Introduction}

{}From time to time, non-Hermitian Hamiltonians have found applications in several
research areas, such as nuclear physics, quantum field theory, condensed matter physics,
and biology. Since the pioneering works of Bessis~\cite{bessis} and Bender and
Boettcher~\cite{bender98}, a subclass of such Hamiltonians, containing operators
invariant under joint action of parity (P: $x \to -x$, $p \to -p$) and time reversal (T: $x
\to x$, $p \to -p$, ${\rm i} \to - {\rm i}$), has become a subject matter of considerable
interest. One of the main reasons for this is that there is a strong analytical and numerical
evidence supporting the conjecture that, except when PT symmetry is spontaneously
broken, the bound-state eigenvalues of these Hamiltonians are real (see
e.g.~\cite{bender98}--\cite{mezincescu}).\par
%
%
PT-symmetric quantum mechanics is often seen as a testing bench for nonconventional
quantum field theories wherein PCT invariance (or PT invariance in scalar theories) plays a
leading role (see e.g.~\cite{bender97, bender01}). This may be compared to the
introduction of supersymmetric quantum mechanics by Witten~\cite{witten} as a testing
ground for non-perturbative methods of achieving supersymmetry breaking in field
theory.\par
%
%
Both the field-theoretic interpretation and the phenomenological relevance of complex
potentials in quantum mechanics make it necessary to better understand some
fundamental issues connected with the replacement of Hermiticity by PT-symmetry, such
as the physical interpretation of the Hamiltonian eigenfunctions~\cite{bender00,
znojil01, japaridze, kretschmer,ahmed}.\par
%
%
The present work is intended as a contribution to such an understanding. Our purpose is
twofold. Firstly, we will generalize to PT-symmetric quantum mechanics the conservation
law of standard quantum mechanics~\cite{schiff} (here formulated for one-dimensional
systems)
\begin{equation}
  \frac{\partial P(x,t)}{\partial t} + \frac{\partial J(x,t)}{\partial x} = 0,
  \label{eq:CE}
\end{equation}
similar to the continuity equation of classical hydrodynamics and relating the change in
time of the position probability density $P(x,t) = |\psi(x,t)|^2$ to the gradient of the
probability current density $J(x,t) = (\hbar/2m {\rm i}) \left[\psi^*(x,t) (\partial
\psi(x,t) / \partial x) - \psi(x,t) (\partial \psi^*(x,t) / \partial x)\right]$. Secondly, we
will discuss with some detailed examples a proposal for modifying the definition of the
normalization condition of bound-state eigenfunctions in accordance with this new
conservation law.\par
%
%
\section{Generalized Continuity Equation and Some of Its Consequences}

Let us start with the time-dependent Schr\"odinger equation for a single particle in 
one dimension,
\begin{equation}
  {\rm i} \hbar \frac{\partial \psi(x,t)}{\partial t} = - \frac{\hbar^2}{2m} 
  \frac{\partial^2 \psi(x,t)}{\partial x^2} + V(x) \psi(x,t),  \label{eq:SE}
\end{equation}
where the potential $V(x)$ is assumed to be complex and PT-symmetric, i.e., 
${\rm PT} V(x) {\rm PT} = V^*(-x) = V(x)$. As a consequence, the function
$\psi^*(-x,t)$ satisfies the equation
\begin{equation}
  - {\rm i} \hbar \frac{\partial \psi^*(-x,t)}{\partial t} = - \frac{\hbar^2}{2m} 
  \frac{\partial^2 \psi^*(-x,t)}{\partial x^2} + V(x) \psi^*(-x,t),  \label{eq:SE*}
\end{equation}
where the potential is the same as in~(\ref{eq:SE}).\par
%
%
Consider now eq.~(\ref{eq:SE}) for some solution $\psi_1(x,t)$ and
eq.~(\ref{eq:SE*}) for some other solution $\psi_2(x,t)$, which may be equal to or
different from $\psi_1(x,t)$. Multiplying (\ref{eq:SE}) by $\psi_2^*(-x,t)$,
(\ref{eq:SE*}) by $\psi_1(x,t)$, and subtracting, we obtain
\begin{equation}
  \frac{\partial}{\partial t} \left[\psi_2^*(-x,t) \psi_1(x,t)\right] + \frac{\hbar}
  {2m {\rm i}} \frac{\partial}{\partial x} \left[\psi_2^*(-x,t) \frac{\partial
  \psi_1(x,t)}{\partial x} - \psi_1(x,t) \frac{\partial \psi_2^*(-x,t)}{\partial x}
  \right] = 0. \label{eq:GCE12} 
\end{equation}
For $\psi_2(x,t) = \psi_1(x,t) = \psi(x,t)$, it is clear that eq.~(\ref{eq:GCE12}) is a
natural generalization of the continuity equation~(\ref{eq:CE}) of standard quantum
mechanics, namely
\begin{equation}
  \frac{\partial P_{PT}(x,t)}{\partial t} + \frac{\partial J_{PT}(x,t)}{\partial x} = 0,
\end{equation}
with
\begin{equation}
  P_{PT}(x,t) = \psi^*(-x,t) \psi(x,t), \quad J_{PT}(x,t) = \frac{\hbar}{2m {\rm i}}
  \left[\psi^*(-x,t) \frac{\partial \psi(x,t)}{\partial x} - \psi(x,t)
  \frac{\partial \psi^*(-x,t)}{\partial x}\right].  
\end{equation}
\par
%
%
If $\psi_1(x,t) \to 0$ and $\psi_2(x,t) \to 0$ for $x \to \pm \infty$, as is normally
required for bound-state wave functions, integration of~(\ref{eq:GCE12}) over the
entire real line leads to the conservation law
\begin{equation}
  \frac{\partial}{\partial t} \int_{-\infty}^{+\infty} dx\, \psi_2^*(-x,t) \psi_1(x,t)
  = 0, \label{eq:cons-norm12}
\end{equation}
or for $\psi_2(x,t) = \psi_1(x,t) = \psi(x,t)$,
\begin{equation}
  \frac{\partial}{\partial t} \int_{-\infty}^{+\infty} dx\, P_{PT}(x,t) = 0.
  \label{eq:cons-norm}
\end{equation}
Note that for Hermitian Hamiltonians, the result corresponding
to~(\ref{eq:cons-norm}) is interpreted in standard quantum mechanics as the
conservation of norm in time~\cite{schiff}.\par
%
%
{}From now on, we are going to restrict ourselves to energy eigenfunctions
\begin{equation}
  \psi_1(x,t) = u_1(x) e^{- \frac{\rm i}{\hbar} E_1 t}, \qquad \psi_2(x,t) = u_2(x)
  e^{- \frac{\rm i}{\hbar} E_2 t}, \label{eq:u-def}
\end{equation}
associated with (possibly complex) energies $E_1$, $E_2$, respectively. On
inserting (\ref{eq:u-def}) into~(\ref{eq:cons-norm12}), we obtain
\begin{equation}
  (E_1 - E_2^*) e^{- \frac{\rm i}{\hbar} (E_1 - E_2^*) t} \int_{-\infty}^{+\infty} dx\,
  u_2^*(-x) u_1(x) = 0.  \label{eq:cons-norm-u} 
\end{equation}
To study the consequences of~(\ref{eq:cons-norm-u}), we have to distinguish
between three cases.\par
%
%
If the two eigenvalues $E_1$ and $E_2$ are real, then $E_2 \ne E_1$ implies the
vanishing of the integral on the left-hand side of~(\ref{eq:cons-norm-u}), namely
\begin{equation}
  \int_{-\infty}^{+\infty} dx\, u_2^*(-x) u_1(x) = 0, \label{eq:R-scalprod} 
\end{equation}
whereas for $E_2 = E_1 = E$, eq.~(\ref{eq:cons-norm-u}) is automatically satisfied.
In such a case, if $E$ is nondegenerate, $u_2(x)$ is proportional to $u_1(x) = u(x)$
and following~(\ref{eq:cons-norm}) we may propose
\begin{equation}
  \int_{-\infty}^{+\infty} dx\, u^*(-x) u(x)  \label{eq:R-norm}
\end{equation}
as a counterpart of the standard normalization integral in PT-symmetric quantum
mechanics. In the same way, the left-hand side of~(\ref{eq:R-scalprod}) can be seen
as a counterpart of the scalar product. It should be noted that in giving up the
Hermiticity condition of the Hamiltonian and replacing it by PT symmetry, some
properties of norms and scalar products have been lost. For instance, it is obvious
that the integral in~(\ref{eq:R-norm}) is not positive-definite and may therefore be
called pseudo-norm~\cite{znojil01}.\par
%
%
If instead only one of the eigenvalues (for instance $E_1$) is real and the other
($E_2$) is complex, eq.~(\ref{eq:R-scalprod}) again follows
from~(\ref{eq:cons-norm-u}).\par
%
%
{}Finally, if both eigenvalues $E_1$ and $E_2$ are complex,
eq.~(\ref{eq:cons-norm-u}) implies condition~(\ref{eq:R-scalprod}) if $E_2 \ne
E_1^*$. This is true in particular for $E_2 = E_1 = E$. In such a case, if $E$ is
nondegenerate, thence $u_2(x)$ is proportional to $u_1(x) = u(x)$ and
condition~(\ref{eq:R-scalprod}) becomes
\begin{equation}
  \int_{-\infty}^{+\infty} dx\, u^*(-x) u(x) = 0.  
\end{equation}
On the contrary, if $E_2 = E_1^* = E^*$, then eq.~(\ref{eq:cons-norm-u}) is
automatically satisfied and we may take
\begin{equation}
  \int_{-\infty}^{+\infty} dx\, u_2^*(-x) u_1(x)  \label{eq:C-norm}
\end{equation}
as a counterpart of the standard normalization integral. Note that
in~(\ref{eq:C-norm}), $u_1(x)$ and $u_2(x)$ correspond to $E$ and $E^*$,
respectively. It is remarkable that in going from real to complex eigenvalues, the
roles of~(\ref{eq:R-scalprod}) and~(\ref{eq:R-norm}) are interchanged. Such
results agree with the analysis carried out in~\cite{znojil01}.\par
%
%
Let us consider in more detail the case of an eigenfunction $u(x)$ corresponding to
some real, nondegenerate eigenvalue $E$. It satisfies the time-independent
Schr\"odinger equation $- (\hbar^2/2m) d^2u(x)/d^2x + V(x) u(x) = E u(x)$, and so is
the case of $u^*(-x)$. Since $E$ is nondegenerate, we must have $u^*(-x) = c u(x)$
for some complex constant $c$. From this relation, it follows that $u^*(x) = c u(-x)
= |c|^2 u^*(x)$, hence $|c| = 1$. We conclude that
\begin{equation}
  u^*(-x) = e^{{\rm i} \phi} u(x), \qquad 0 \le \phi < 2 \pi.  \label{eq:u*}
\end{equation}
\par
%
%
If $\phi = 0$ (resp.\ $\phi = \pi$) in~(\ref{eq:u*}), then $u^*(-x) = u(x)$ (resp.\
$u^*(-x) = - u(x)$) or, in other words, $u(x)$ is PT-symmetric (resp.\
PT-antisymmetric). If however $\phi \ne 0, \pi$, it is always possible to convert
$u(x)$ into such a function by multiplying it by some appropriate phase factor. The
functions
\begin{equation}
  v_{\sigma}(x) \equiv e^{\frac{\rm i}{2}\left(\phi - \frac{\pi}{2} + \sigma 
  \frac{\pi}{2}\right)} u(x), \qquad \sigma = \pm 1, \label{eq:v}
\end{equation}
indeed satisfy the relations $v_{\sigma}^*(-x) = \sigma v_{\sigma}(x)$, $\sigma = \pm
1$ and are PT-symmetric ($\sigma = +1$) or PT-antisymmetric ($\sigma = -1$).\par
%
%
We propose to normalize $u(x)$, or the corresponding $v_{\sigma}(x)$, according to
the prescription
\begin{equation}
  \int_{-\infty}^{+\infty} dx\, u^*(-x) u(x) = \sigma \int_{-\infty}^{+\infty} dx\,
  [v_{\sigma}(x)]^2 = \sigma.  \label{eq:norm}
\end{equation}
In the following, we illustrate the use of the above condition by considering some specific
cases of PT-symmetric model potentials, namely, PT-symmetric oscillator,
generalized Poschl-Teller, and Scarf II potentials. In particular, we show
that for this rule to be meaningful, we have to tell which PT-parity $\sigma$, i.e.,
which function $v_{\sigma}(x)$, we are going to associate to any given
eigenfunction $u(x)$ with real eigenvalue.\par
%
%
\section{Modified Normalization for the PT-Symmetric Oscillator}

The potential for the PT-symmetric oscillator is given by~\cite{znojil99}
\begin{equation}
  V^{(\alpha)}(x) = (x - {\rm i} c)^2 + \frac{\alpha^2 - \frac{1}{4}}{(x - {\rm i} c)^2}, 
\end{equation}
where $\alpha > 0$. It can be obtained from the usual three-dimensional
radial harmonic oscillator potential by effecting a complex shift of coordinate $x \to x -
{\rm i} c$, $c > 0$, and replacing $l$ by $\alpha - \frac{1}{2}$. Although it is beset
with a centrifugal-like core, the shift of the singularity off the integration path makes the
corresponding Hamiltonian exactly solvable on the entire real line for any $\alpha > 0$, like
the linear harmonic oscillator to which it reduces for $\alpha = \frac{1}{2}$ and
$c=0$.\par
%
%
If $\alpha$ is different from an integer (which we shall assume here), the PT-symmetric
oscillator Hamiltonian has a double series of energy eigenvalues $E^{(\alpha)}_{qn}$,
which may be distinguished by a quantum number $q = \pm 1$~\cite{znojil99}. For $\hbar
= 2m = 1$, they are given by
\begin{equation}
  E^{(\alpha)}_{qn} = 4n + 2 - 2q \alpha, \qquad n = 0, 1, 2, \ldots. \label{eq:HO-E}
\end{equation}
The corresponding eigenfunctions are expressible in terms of generalized Laguerre
polynomials:
\begin{equation}
  u^{(\alpha)}_{qn}(x) = {\cal N}^{(\alpha)}_{qn} e^{-\frac{1}{2}(x - {\rm i} c)^2} 
  (x - {\rm i} c)^{-q\alpha + \frac{1}{2}} L_n^{(-q \alpha)}[(x - {\rm i} c)^2].
  \label{eq:HO-u}
\end{equation}
Here ${\cal N}^{(\alpha)}_{qn}$ is some yet undetermined normalization coefficient.
Although the Laguerre polynomials in~(\ref{eq:HO-u}) are in principle defined only for $-q
\alpha > -1$, i.e., for $\alpha$ values in the interval $0 < \alpha < 1$, their definition can
be extended to any values of $\alpha$. In the interval $0 < \alpha < 1$
(and only in it), however, the energies~(\ref{eq:HO-E}) are ordered according to
increasing values of $N = 2n + \frac{1}{2} (1-q) = 0$, 1, 2,~\ldots, and alternatively
correspond to $q = +1$ and $q = -1$. In the special case of the linear harmonic oscillator
($\alpha = \frac{1}{2}$ and $c=0$), $q$ becomes the parity of the eigenfunctions. For
this reason, $q$ is called quasi-parity of the PT-symmetric oscillator.\par  
%
%
It is straightforward to see that $\left[u^{(\alpha)}_{qn}(-x)\right]^*$ is related to
$u^{(\alpha)}_{qn}(x)$ as shown in~(\ref{eq:u*}), where the phase is given by $\phi =
\pi (- q\alpha + \frac{1}{2}) - 2\nu$ (up to a multiple of $2\pi$). Here $\nu$ denotes
the phase of the normalization coefficient ${\cal N}^{(\alpha)}_{qn} = |{\cal
N}^{(\alpha)}_{qn}| e^{{\rm i} \nu}$.\par
%
%
We now choose to identify the PT-parity $\sigma$ of the functions $v_{\sigma}(x)$
in~(\ref{eq:v}) with the quasi-parity $q$ of the eigenfunctions~(\ref{eq:HO-u}). As a
consequence, the functions
\begin{equation}
  v^{(\alpha)}_{qn}(x) = e^{{\rm i} q \frac{\pi}{2} (-\alpha+\frac{1}{2})} \left|{\cal
  N}^{(\alpha)}_{qn}\right| e^{-\frac{1}{2}(x - {\rm i} c)^2} 
  (x - {\rm i} c)^{-q\alpha + \frac{1}{2}} L_n^{(-q \alpha)}[(x - {\rm i} c)^2]
  \label{eq:HO-v}
\end{equation}
are PT-symmetric or PT-antisymmetric for $q = +1$ or $q = -1$, respectively.\par
%
%
Let us normalize the functions $u^{(\alpha)}_{qn}(x)$ and $v^{(\alpha)}_{qn}(x)$
according to the rule~(\ref{eq:norm}), where $\sigma$ is replaced by $q$.
From~(\ref{eq:HO-u}), we obtain
\begin{equation}
  \int_{-\infty}^{+\infty} dx\, \left[u^{(\alpha)}_{qn}(-x)\right]^* u^{(\alpha)}_{qn}(x)
  = e^{{\rm i} \pi (- q\alpha + \frac{1}{2})} \left|{\cal N}^{(\alpha)}_{qn}\right|^2 {\cal
  I}^{(\alpha)}_{qn}, \label{eq:HO-I} 
\end{equation}
where
\begin{equation}
  {\cal I}^{(\alpha)}_{qn} = \int_{-\infty}^{+\infty} dx\, e^{-(x-{\rm i}c)^2} (x-{\rm i}c)^
  {- 2q\alpha + 1} \left\{L^{(-q\alpha)}_n[(x-{\rm i}c)^2]\right\}^2.   
\end{equation}
\par
%
%
To calculate ${\cal I}^{(\alpha)}_{qn}$, we consider the function of the complex
variable $z$ given by $f(z) = e^{-z^2} z^{- 2q\alpha + 1}
\left[L^{(-q\alpha)}_n(z^2)\right]^2$. It is analytic everywhere except at $z=0$, which
is a branch point. Let us take the negative real axis as the branch cut. As the integral
${\cal I}^{(\alpha)}_{qn}$ can be seen as the limit for $R \to \infty$ of
$\int_{-R}^{+R}dz\, f(z)$ for $z = x - {\rm i}c$ (i.e., the integral on segment AB in
Fig.~1), we close a contour $\Gamma \equiv {\rm ABCDEF}$  in the complex plane,
avoiding the singularity by using a semi-circle ${\rm C}_{\rho}$ of radius $\rho$. Since
$f(z)$ is analytic within and on $\Gamma$, by Cauchy's theorem, we have
$\int_{\Gamma} dz\, f(z) = 0$. Evaluating this integral and taking the limits $R \to
\infty$ and $\rho \to 0$, we obtain that ${\cal I}^{(\alpha)}_{qn}$ converges if $q
\alpha < 1$, which is consistent with $q = \pm 1$ and $0 < \alpha < 1$.\par
%
%
As a result, eq.~(\ref{eq:HO-I}) becomes
\begin{equation}
  \int_{-\infty}^{+\infty} dx\, \left[u^{(\alpha)}_{qn}(-x)\right]^* u^{(\alpha)}_{qn}(x)
  = \left|{\cal N}^{(\alpha)}_{qn}\right|^2 \cos \pi(-q\alpha + \case{1}{2})
  \frac{\Gamma(- q\alpha + n +1)}{n!}, \label{eq:HO-norm} 
\end{equation}
if $0 < \alpha < 1$. Since the cosine on the right-hand side of~(\ref{eq:HO-norm}) is
positive or negative according to whether $q = +1$ or $q = -1$, the result
in~(\ref{eq:HO-norm}) is consistent with condition~(\ref{eq:norm}). The latter
therefore leads to
\begin{equation}
  \left|{\cal N}^{(\alpha)}_{qn}\right| = \left(\frac{n!}{\Gamma(- q\alpha + n +1) \cos \pi
  ( -\alpha + \case{1}{2})}\right)^{1/2}. \label{eq:HO-N}
\end{equation}
As compared with the standard harmonic oscillator, the presence of a cosine on the
right-hand side of~(\ref{eq:HO-N}) is a new feature. Such a cosine disappears for $\alpha
= \frac{1}{2}$, so eq.~(\ref{eq:HO-N}) agrees with the known normalization coefficients
of the linear oscillator ($\alpha = \frac{1}{2}$, $q = \pm 1$) and of the
three-dimensional radial oscillator ($\alpha = \frac{1}{2}$, $l=0$, $q = -1$).\par
%
%
\section{Modified Normalization for the PT-Symmetric Generalized P\"oschl-Teller
Potential}

The PT-symmetric generalized P\"oschl-Teller potential is given by~\cite{bagchi00c}
\begin{equation}
  V^{(A,B)}(x) = \left[B^2 + A (A+1)\right] \cosech^2 \tau - B (2A+1) \cosech
  \tau \coth \tau, \qquad \tau \equiv x - {\rm i} \gamma, \label{eq:PT-V} 
\end{equation}
where $B > A + \frac{1}{2} > 0$ and $- \frac{\pi}{4} \le \gamma < 0$ or $0 < \gamma <
\frac{\pi}{4}$. By using  an sl(2,\C) algebraic framework, it was recently
shown~\cite{bagchi00c} that whenever $B - A - \frac{1}{2}$ is different from an integer, 
the corresponding Hamiltonian has two series of real energy eigenvalues
$E^{(A,B)}_{qn}$, which may be distinguished by a label $q = \pm 1$ (here called
quasi-parity for convenience sake) and which for $\hbar = 2m = 1$ are given  by 
\begin{eqnarray}
  E^{(A,B)}_{+n} & = & - \left(B - \case{1}{2} - n\right)^2, \qquad n = 0, 1, \ldots,
        n_{+ max}, \nonumber \\
  && B - \case{3}{2} \le n_{+ max} < B - \case{1}{2}, \\
  E^{(A,B)}_{-n} & = & - \left(A - n\right)^2, \qquad n = 0, 1, \ldots, n_{- max},
        \nonumber \\
  && A - 1 \le n_{- max} < A,
\end{eqnarray}
where $B > \frac{1}{2}$ and $A > 0$. The accompanying eigenfunctions read
\begin{eqnarray}
  u^{(A,B)}_{qn}(x) & = & {\cal N}^{(A,B)}_{qn} (y-1)^{\frac{1}{2}(\lambda +
        \frac{1}{2})} (y+1)^{\frac{1}{2}(\mu + \frac{1}{2})} P_n^{(\lambda, \mu)}(y) 
        \nonumber \\
  & = & {\cal N}^{(A,B)}_{qn} 2^{\frac{1}{2}(\lambda + \mu + 1)}
        \left(\sinh\case{\tau}{2}\right)^{\lambda + \frac{1}{2}}
        \left(\cosh\case{\tau}{2}\right)^{\mu +
        \frac{1}{2}} P_n^{(\lambda, \mu)}(\cosh \tau),
\end{eqnarray}
where $\lambda = q (A - B + \frac{1}{2})$, $\mu = - A - B - \frac{1}{2}$,  $y = \cosh
\tau$, $P^{(\lambda, \mu)}_n(y)$ is a Jacobi polynomial, and ${\cal
N}^{(A,B)}_{qn}$ some normalization coefficient.\par
%
%
In the limit $\gamma \to 0$, the real generalized P\"oschl-Teller potential is recovered
and only the eigenvalues $E^{(A,B)}_{-n}$ and corresponding eigenfunctions
$u^{(A,B)}_{-n}(x)$ survive~\cite{cooper}. Since this potential is singular, it has to be
restricted to the half-line $(0, + \infty)$. The complexified potential, as given above, gets
regularized on performing the shift $x \to x - {\rm i} \gamma$ and so may be considered
on the entire real line, as it was the case for the PT-symmetric harmonic oscillator in the
previous section.\par
%
%
The analogy between these two PT-symmetric potentials also extends to the
normalization problem. The phase $\phi$ in~(\ref{eq:u*}) is now given (up to a multiple of
$2\pi$) by $\phi = \pi (\lambda + \frac{1}{2}) - 2 \nu$, where $\nu$ denotes the
phase of the normalization coefficient. This shows that the role of $\alpha$ for the
PT-symmetric oscillator is now played by the constant $-q \lambda = B - A -
\frac{1}{2}$. As in sect.~3, the PT-parity $\sigma$ may be identified with $q$, leading
to PT-symmetric and PT-antisymmetric eigenfunctions $v^{(A,B)}_{qn}(x)$ ($q =
\pm1$), given by
\begin{equation}
  v^{(A,B)}_{qn}(x) = e^{{\rm i} \frac{\pi}{2} (\lambda + \frac{1}{2} q)} \left|{\cal
  N}^{(A,B)}_{qn}\right| 2^{\frac{1}{2}(\lambda + \mu + 1)}
  \left(\sinh\case{\tau}{2}\right)^{\lambda + \frac{1}{2}}
  \left(\cosh\case{\tau}{2}\right)^{\mu + \frac{1}{2}} P_n^{(\lambda, \mu)}
  (\cosh \tau).
\end{equation}
Applying again prescription~(\ref{eq:norm}) to normalize the eigenfunctions, we are led to
the integral of the function $f(z) =  (\sinh\frac{z}{2})^{2\lambda +1}
(\cosh\frac{z}{2})^{2\mu +1} \left[P_n^{(\lambda, \mu)}(\cosh z)\right]^2$ on the
contour $\Gamma$ of Fig.~1, where $c$ is replaced by $\gamma$ (and the latter is
assumed positive for simplicity's sake). In this respect, it is worth noting that among the
branch points of $f(z)$ at $z=0$, $\pm {\rm i} \pi$, $\pm 2{\rm i} \pi$,
$\pm 3{\rm i} \pi$,~\ldots, we only have to take care of the first one because of the
range of $\gamma$, as stated below eq.~(\ref{eq:PT-V}).\par
%
%
The final result for the normalization integral reads
\begin{equation}
  \int_{-\infty}^{+\infty} dx\, \left[u^{(A,B)}_{qn}(-x)\right]^* u^{(A,B)}_{qn}(x)
  = 2 \left|{\cal N}^{(A,B)}_{qn}\right|^2 \cos \left[\pi (\lambda + \case{1}{2})\right]
  I^{(\lambda, \mu)}_{n}, \label{eq:PT-norm}
\end{equation}
provided $\lambda$ is restricted by the condition $\lambda > -1$, which is consistent with
$q = \pm 1$ and $A + \frac{1}{2} < B < A + \frac{3}{2}$. In~(\ref{eq:PT-norm}), 
$I^{(\lambda,\mu)}_{n}$ is the real integral
\begin{equation}
  I^{(\lambda,\mu)}_{n} = \int_1^{\infty} dt\, (t-1)^{\lambda} (t+1)^{\mu}
  \left[P^{(\lambda, \mu)}_n(t)\right]^2,
\end{equation}
which, for $q = -1$, appears in the calculation of the normalization coefficient of the real
potential eigenfunctions. It is convergent and positive for the above-mentioned restricted
range of parameters and both values of $q$.  As far as we know, no closed analytical
formula is known for it, except when $n=0$. In such a case, the change of variable $s =
1/t$ leads to~\cite{gradshteyn}
\begin{equation}
  I^{(\lambda,\mu)}_0 = 2^{\lambda+\mu+1} \frac{\Gamma(-\lambda-\mu-1) 
  \Gamma(\lambda+1)}{\Gamma(-\mu)}.
\end{equation}
Furthermore, the cosine on the right-hand side
of~(\ref{eq:PT-norm}) is positive or negative according to whether $q = +1$ or $q = -1$.
Hence eq.~(\ref{eq:PT-norm}) is consistent with~(\ref{eq:norm}), from which it follows
that
\begin{equation}
  \left|{\cal N}^{(A,B)}_{qn}\right| = \left\{2 \cos [\pi (A-B+1)]
  I^{(\lambda,\mu)}_{n}\right\}^{-1/2}.
  \label{eq:PT-N}
\end{equation}
The cosine disappears from~(\ref{eq:PT-N}) if $B = A+1$. For this choice of parameters, 
potential~(\ref{eq:PT-V}) reduces to $V^{(A,A+1)} = - \frac{1}{2} (A+1) (2A+1)
\sech^2 \frac{\tau}{2}$, which remains nonsingular in the limit $\gamma \to 0$.\par
%
%
As a final point, it is worth mentioning that the generalized P\"oschl-Teller
potential~(\ref{eq:PT-V}) is related to the PT-symmetric P\"oschl-Teller II potential
considered in~\cite{znojil00} through a complex point canonical coordinate
transformation~\cite{bagchi00c}, which also connects the corresponding eigenfunctions.
The results of the present section can therefore easily be applied to the latter
potential.\par
%
%
\section{Modified Normalization for the PT-Symmetric Scarf II Potential}

The PT-symmetric Scarf II potential is defined by~\cite{bagchi00c}
\begin{equation}
  V^{(A,B)}(x) = - \left[B^2 +A (A+1)\right] \sech^2 x + {\rm i} B (2A+1) \sech x \tanh
x,
\end{equation}
where $A > B - \case{1}{2} > 0$. It was shown using sl(2,\C) as a
tool~\cite{bagchi00c} that for values of $A - B + \frac{1}{2}$ different from an
integer, the corresponding Hamiltonian has a double series of energy eigenvalues
$E^{(A,B)}_{qn}$, distinguished by a label $q = \pm1$ (again called quasi-parity for
convenience sake), and given by   
\begin{eqnarray}
  E^{(A,B)}_{+n} & = & - \left(A - n\right)^2, \qquad n = 0, 1, \ldots,
        n_{+ max}, \nonumber \\
  && A - 1 \le n_{+ max} < A,  \\
  E^{(A,B)}_{-n} & = & - \left(B - \case{1}{2} - n\right)^2, \qquad n = 0, 1, \ldots, 
        n_{-max}, \nonumber \\
  && B - \case{3}{2} \le n_{- max} < B - \case{1}{2}, \label{eq:Scarf-E}
\end{eqnarray}
if $\hbar = 2m = 1$, $A > 0$, and $B > \frac{1}{2}$. The accompanying eigenfunctions
read
\begin{equation}
  u^{(A,B)}_{qn}(x) = {\cal N}^{(A,B)}_{qn} (\sech x)^{-\frac{1}{2}(\lambda + \mu + 1)}
  e^{- \case{\rm i}{2}(\lambda - \mu) \arctan(\sinh x)}
  P_n^{(\lambda,\mu)}({\rm i} \sinh x) \label{eq:Scarf-u}
\end{equation}
in terms of Jacobi polynomials. Here $\lambda = q (- A + B - \frac{1}{2})$, $\mu
= - A - B - \frac{1}{2}$, and ${\cal N}^{(A,B)}_{qn}$ is some normalization
coefficient.\par
%
%
Contrary to the potentials considered in the two previous sections, the PT-symmetric
Scarf II potential is not obtained from the corresponding real potential~\cite{cooper} by a
complex shift of coordinate, but instead by complexifying one parameter ($B \to {\rm i}
B$). Such a complexification is responsible for the appearance of the additional
series~(\ref{eq:Scarf-E}) of energy eigenvalues.\par
%
%
Comparing (\ref{eq:Scarf-u}) with~(\ref{eq:u*}), we find (up to a multiple of $2\pi$) $\phi
= - 2 \nu$, where $\nu$ is the phase of the normalization coefficient. If we identify
PT-parity with quasi-parity again, we obtain
\begin{equation}
  v^{(A,B)}_{qn}(x) = e^{{\rm i} (q-1) \frac{\pi}{4}} \left|{\cal N}^{(A,B)}_{qn}\right|
  (\sech x)^{-\frac{1}{2}(\lambda + \mu + 1)} e^{- \frac{\rm i}{2}(\lambda - \mu)
  \arctan(\sinh x)} P_n^{(\lambda,\mu)}({\rm i} \sinh x),
\end{equation}
where $\exp[{\rm i} (q-1) \frac{\pi}{4}] = 1$ for PT-symmetric functions and $- \rm i$
for PT-antisymmetric ones.\par
%
%
The normalization integral is now
\begin{eqnarray}
  \lefteqn{\int_{-\infty}^{+\infty} dx\, \left[u^{(A,B)}_{qn}(-x)\right]^* 
         u^{(A,B)}_{qn}(x)} \nonumber \\
  & = & \left|{\cal N}^{(A,B)}_{qn}\right|^2 \int_{-\infty}^{+\infty} dx\, (\sech x)^{-
         \lambda - \mu -1} e^{- {\rm i} (\lambda - \mu) \arctan(\sinh x)}
         \left[P_n^{(\lambda, \mu)}({\rm i} \sinh x)\right]^2. \label{eq:Scarf-norm}
\end{eqnarray}
\par
%
%
Let us consider, for instance, the $n=0$ case corresponding to the lowest states of both
series of energy levels. The integral on the right-hand side of~(\ref{eq:Scarf-norm}) is
then easily calculated by performing the change of variable $\tan y = \sinh x$. The results
read~\cite{gradshteyn}
\begin{eqnarray}
  \int_{-\infty}^{+\infty} dx\, \left[u^{(A,B)}_{+0}(-x)\right]^* u^{(A,B)}_{+0}(x)
        & = & \left|{\cal N}^{(A,B)}_{+0}\right|^2 \frac{\pi \Gamma(2A)}{2^{2A-1}
        \Gamma(A - B + \frac{1}{2}) \Gamma(A + B + \frac{1}{2})}, \label{eq:Scarf-norm+}
        \\
  \int_{-\infty}^{+\infty} dx\, \left[u^{(A,B)}_{-0}(-x)\right]^* u^{(A,B)}_{-0}(x)
        & = & \left|{\cal N}^{(A,B)}_{-0}\right|^2 \frac{\pi \Gamma(2B-1)}{2^{2B-2}
        \Gamma(B - A - \frac{1}{2}) \Gamma(B + A + \frac{1}{2})}. \label{eq:Scarf-norm-}
\end{eqnarray}
\par
%
%
The right-hand side of~(\ref{eq:Scarf-norm+}) is positive for any allowed values of $A$
and $B$ because the arguments of all three gamma functions are positive. This is
compatible with the normalization condition~(\ref{eq:norm}) with $\sigma = q = +1$. On
the right-hand side of~(\ref{eq:Scarf-norm-}), on the contrary, the argument of
$\Gamma(B - A - \frac{1}{2})$ is negative while those of the two remaining gamma
functions are positive, which means that the sign of the normalization integral for $\sigma
= q = -1$ depends on the relative magnitudes of $A$ and $B - \frac{1}{2}$. A negative
sign compatible with~(\ref{eq:norm}) is obtained for $B -
\frac{1}{2} + 2k < A < B + \frac{1}{2} + 2k$, $k=0$, 1, 2,~\ldots. In particular, for
$k=0$, we get a condition $B - \frac{1}{2} < A < B + \frac{1}{2}$ rather similar to the
conditions previously encountered for the PT-symmetric oscillator and generalized
P\"oschl-Teller potentials.\par
%
%
\section{Conclusion}

To conclude, we have derived, in the set up of a very general framework,
the conservation law pertaining to PT-symmetric quantum mechanical systems.
We have found, as a consequence, that the normalization integral ceases
to be always positive definite. The possibilty of encountering non-positive
definite norms is illustrated by means of some specific examples of 
PT-symmetric potentials.
%
%
\section*{Acknowledgments}

B.\ B.\ thanks Professor Kalipada Das and Professor Mithil Ranjan Gupta for their
enlightening comments. C.\ Q.\ is a Research Director of the National Fund for Scientific
Research (FNRS), Belgium. Participation of M.\ Z.\ in this work was supported by the GA AS
grant Nr.\ A 104 8004.\par
%
%
 
%
%
\newpage
\section*{Figure caption}

{}Fig.\ 1. Contour $\Gamma$ in the complex plane.
%
%
\newpage
\begin{picture}(160,120)(-75,-80)

\put(-70,0){\vector(1,0){140}}
\put(0,-40){\vector(0,1){55}}

\put(-60,0){\line(0,1){2}}
\put(60,0){\line(0,1){2}}
\put(-5,0){\line(0,1){2}}
\put(5,0){\line(0,1){2}}

\put(-60,-30){\thicklines\line(1,0){120}}
\put(60,-30){\thicklines\line(0,1){29}}
\put(5,-1){\thicklines\line(1,0){55}}
\put(0,-1){\thicklines\oval(10,10)[b]}
\put(-60,-1){\thicklines\line(1,0){55}}
\put(-60,-30){\thicklines\line(0,1){29}}

\put(30,-30){\thicklines\vector(1,0){0}}
\put(7,-15){\vector(-1,2){4.5}}

\put(77,0){\makebox(0,0){\Large ${\rm Re\ }z$}}
\put(0,20){\makebox(0,0){\Large ${\rm Im\ }z$}}
\put(28,-34){\makebox(0,0){\Large $\Gamma$}}
\put(-60,-34){\makebox(0,0){\Large A}}
\put(60,-34){\makebox(0,0){\Large B}}
\put(63,-3){\makebox(0,0){\Large C}}
\put(8,-4){\makebox(0,0){\Large D}}
\put(-8,-4){\makebox(0,0){\Large E}}
\put(-63,-3){\makebox(0,0){\Large F}}
\put(10,-18){\makebox(0,0){\Large ${\rm C}_{\rho}$}}
\put(-60,5){\makebox(0,0){\Large $-R$}}
\put(-5,5){\makebox(0,0){\Large $- \rho$}}
\put(5,5){\makebox(0,0){\Large $+ \rho$}}
\put(60,5){\makebox(0,0){\Large $+R$}}
\put(-5,-33){\makebox(0,0){\Large $-c$}}

\end{picture}

\centerline{Figure 1}    
 
\end{document}